\documentclass[3p,a4paper,12pt]{elsarticle}

\usepackage{amssymb}
\usepackage{amsmath}
\usepackage{cancel}
\usepackage{hyperref}
\hypersetup{colorlinks=true, citecolor=blue, linkcolor=blue, urlcolor=blue}

\begin{document}

\begin{frontmatter}



\title{An improved nonlocal electron heat transport model for magnetized plasmas}


\author[1]{Z. H. Chen} 
\author[1]{Z. Q. Zhao} 
\author[1]{X. H. Yang\corref{cor1}} 
\ead{xhyang@nudt.edu.cn}
\author[1]{L. R. Li} 
\author[1]{B. Zeng} 
\author[1]{Z. Li} 
\author[1]{B. H. Xu}
\author[1]{G. B. Zhang} 
\author[1]{H. H. Ma} 
\author[2]{M. Tang}
\author[3,4]{Y. Y. Ma} 
\author[3,4]{H. Xu} 
\author[1]{F. Q. Shao} 
\author[5,6]{J. Zhang} 

\cortext[cor1]{Corresponding author}

\affiliation[1]{organization={College of Science},
            addressline={National University of Defense Technology}, 
            city={Changsha},
            citysep={},
            postcode={410073}, 
            country={China}}

\affiliation[2]{organization={School of Mathematics, Institute of Natural Sciences and MOE-LSC},
	addressline={Shanghai Jiao Tong University}, 
	city={Shanghai},
	citysep={},
	postcode={200240}, 
	country={China}}
	
\affiliation[3]{organization={School of Automation and Electronic Information},
	addressline={Xiangtan University}, 
	city={Xiangtan},
	citysep={},
	postcode={411105}, 
	country={China}}
	
\affiliation[4]{organization={School of Physics and Electronics},
	addressline={Hunan University}, 
	city={Changsha},
	citysep={},
	postcode={410082}, 
	country={China}}

\affiliation[5]{organization={Collaborative Innovation Centre of IFSA},
	addressline={Shanghai Jiao Tong University}, 
	city={Shanghai},
	citysep={},
	postcode={200240}, 
	country={China}}
	
\affiliation[6]{organization={Key Laboratory for Laser Plasmas (Ministry of Education), School of Physics and Astronomy},
	addressline={Shanghai Jiao Tong University}, 
	city={Shanghai},
	citysep={},
	postcode={200240}, 
	country={China}}

\begin{abstract}
Distortions in the electron distribution function driven by intense temperature gradients critically influence the generation and evolution of heat flux and magnetic fields in plasmas under the condition of inertial confinement fusion. Describing such kinetic behaviors at large spatiotemporal scales typically requires multigroup models based on simplified Vlasov-Fokker-Planck equations. However, the accuracy of existing multigroup models remains uncertain, without a well-defined methodology for implementing nonlocal magnetic field corrections. This paper develops an improved nonlocal multigroup model for magnetized plasmas. The advancements comprise: (i) a revised source term in the diffusion equations, (ii) a Biermann-producing electric field equation incorporating the density perturbation, and (iii) a nonlocal correction method for the Nernst velocity. The numerical method for the anisotropic heat conduction equation is analyzed, and three test cases demonstrate that the model accurately predicts the key phenomena arising from nonlocal effects in magnetized plasmas.
\end{abstract}




\end{frontmatter}



\section{Introduction}
\label{sec1}
In laser-driven inertial confinement fusion (ICF), the intense laser heating can generate steep temperature gradients in plasma, thereby causing a distortion of the electron distribution function that deviates from the classical Maxwellian distribution \cite{lucianiNonlocalHeatTransport1983,epperleinTwoDimensionalNonlocalElectron1988,gregoriEffectNonlocalTransport2004,schurtzRevisitingNonlocalElectronEnergy2007,henchenObservationNonlocalHeat2018}. This non-equilibrium state is primarily manifested through two key physical processes. The first is the electron nonlocal thermal conduction, which produces dual effects, i.e., heat flux limiting and preheating. These effects significantly influence the laser energy absorption, ablation velocity, shock strength, target acceleration, implosion compression, and hydrodynamic instabilities \cite{smalyukRayleighTaylorGrowthStabilization2008,huValidationThermalTransportModeling2008,marocchinoComparisonNonlocalHydrodynamic2013,maThermalTransportModeling2022,chenEffectNonlocalTransport2023,chenRoleNonlocalHeat2024,arranAccuracyMultigroupModels2025}. The second is the nonlocal magnetic fields, encompassing nonlocal modifications to the Biermann magnetic field and the nonlocal Nernst effect \cite{khoNonlinearKineticTransport1985,lucianiMagneticFieldNonlocal1985,lanciaTopologyMegagaussMagnetic2014,gaoPrecisionMappingLaserDriven2015}. The Biermann battery mechanism is a fundamental process for generating self-generated magnetic fields in plasmas. In the classical case, noncollinear electron temperature and density gradients drive magnetic field growth. Under nonlocal conditions, however, the growth of the Biermann magnetic field is suppressed \cite{sherlockSuppressionBiermannBattery2020,campbellMeasuringMagneticFlux2022,daviesNonlocalSuppressionBiermann2023,griff-mcmahonMeasurementsExtendedMagnetic2024a}, while a magnetic field can still be generated in the regions devoid of density gradients \cite{kinghamNonlocalMagneticFieldGeneration2002,kinghamImplicitVlasovFokker2004}. Regarding the Nernst effect, it has been shown that under nonlocal scenarios, the electric field peak shifts towards colder regions, giving rise to a characteristic `pre-Nernst' topological profile \cite{brodrickIncorporatingKineticEffects2018,hillEnhancementPressurePerturbations2018a}.

These two physical processes exhibit mutual influence and tight coupling. Firstly, magnetic fields (including both self-generated and externally applied configurations) significantly alter electron transport characteristics, thus modulating electron nonlocal heat conduction. Secondly, the nonlocal heat conduction of electrons changes temperature profiles, which in turn affects the extent of distribution function distortion as well as the generation and evolution of self-generated magnetic field.

The flux limiters were originally conceived to address the suppression of heat flux induced by electron nonlocal heat conduction. Subsequent extensions incorporated corrections for the Nernst velocity \cite{khoNonlinearKineticTransport1985,daviesImportanceElectrothermalTerms2015,walshKineticCorrectionsHeatflow2024}. By combining flux limiters with a nonlocal suppression model for the Biermann magnetic field, such as those proposed by Sherlock et al. \cite{sherlockSuppressionBiermannBattery2020} or Davies et al. \cite{daviesNonlocalSuppressionBiermann2023}, this integrated methodology provides the simplest and fastest method for predicting nonlocal effects in magnetized plasmas. However, this composite model is not entirely accurate. Beyond empirically selected flux limiters, it also fails to comprehensively capture various nonlocal phenomena such as the preheating, the Biermann magnetic field without density gradients, and the enhanced Nernst velocity inside the target.

Accurate modeling of nonlocal effects typically requires Vlasov-Fokker-Planck (VFP) codes. The VFP equations provide a complete microscopic description of electron kinetics, enabling direct calculations of macroscopic properties of interest through velocity moments. However, solving such precise models demands substantial computational time and memory resources. Therefore, based on the simplified VFP equations, some alternative approaches have been proposed to accurately resolve nonlocal electron transport. The most established model is the Schurtz-Nicolaï-Busquet (SNB) multigroup diffusion model \cite{schurtzNonlocalElectronConduction2000,caoImprovedNonlocalElectron2015}, whose accuracy in unmagnetized plasmas aligns well with experiments and kinetic simulations. The SNB model has also been extended to magnetized plasmas \cite{nicolaiPracticalNonlocalModel2006}. Despite the considerable research potential of the magnetized SNB model (hereafter called mSNB model), its complexity has limited its investigation.

In this work, we have implemented refinements, extensions, and calibrations to the original mSNB model. The improved model self-consistently and accurately captures nonlocal effects in magnetized plasmas, including precise prediction of heat flux limiting, preheating, and nonlocal Biermann battery and Nernst effect. Section \ref{sec2} introduces the kinetic equations for the nonlocal heat conduction of magnetized electrons, presenting novel insights into source terms of the diffusion equations, nonlocal suppression of the Biermann magnetic field, and nonlocal Nernst effect. In Section \ref{sec3}, we describe the model implementation and present a stability analysis for the anisotropic heat conduction equation. In Section \ref{sec4}, we tested the model through three basic cases that collectively encompass all significant nonlocal effects, aiming to verify the accuracy of the model. Section \ref{sec5} discusses the test results, model limitations, and potential future work, while Section \ref{sec6} presents our conclusions.

\section{Improved nonlocal transport model for magnetized plasmas}
\label{sec2}
\subsection{The original model}
\label{subsec2.1}

The magnetized electron VFP kinetic equation is
\begin{equation}\label{Eq1}
	\frac{{\partial {f_e}}}{{\partial t}} + \vec v \cdot \nabla {f_e} - \frac{e}{{{m_e}}}\left( {\vec E + \frac{{\vec v}}{c} \times \vec B} \right)\frac{{\partial {f_e}}}{{\partial v}} = {C_{ee}} + {C_{ei}},
\end{equation}

where ${f_e}$ is the electron distribution function, $t$ is time, $\vec v$ is the electron velocity, $-e$ is the elementary charge, ${m_e}$ is the electron mass, $\vec E$ is the electric field, $\vec B$ is the magnetic field, $c$ is the speed of light. ${C_{ee}}$ and ${C_{ei}}$ are electron-electron ($e$-$e$) and electron-ion ($e$-$i$) collision operators, respectively. Direct numerical solution of the complete VFP equation incurs excessive computational cost and therefore needs to be simplified. Assuming the distribution function is nearly isotropic, the spherical harmonics expansion of the distribution function can be truncated to the first order
\begin{equation}\label{Eq2}
	{f_e}\left( {\vec x,\vec v,t} \right) = \frac{{{f_0}}}{{4\pi }} + \frac{{3\vec \Omega   \cdot {{\vec f}_1}}}{{4\pi }},
\end{equation}
where ${f_0}$ is the isotropic part of the distribution function, ${\vec f_1}$ is the anisotropic part, and $\vec \Omega   = \vec v/v$ is the unit vector in the direction of velocity. This first order correction is generally considered sufficiently accurate for describing laser-ablated plasmas \cite{marocchinoComparisonNonlocalHydrodynamic2013,delsorboReducedEntropicModel2015}. Substituting the first order expanded distribution function in Eq. \eqref{Eq2} into the electron kinetic equation \eqref{Eq1} and considering steady-state conditions yields
\begin{equation}\label{Eq3}
	v\nabla  \cdot {\vec f_1} - \frac{{e\vec E}}{{{m_e}}}\frac{1}{{{v^2}}}\frac{\partial }{{\partial v}}\left( {{v^2}{{\vec f}_1}} \right) = {C_0}\left( {{f_0}} \right),
\end{equation}
\begin{equation}\label{Eq4}
	\frac{v}{3}\nabla {f_0} - \frac{{e\vec E}}{{3{m_e}}}\frac{{\partial {f_0}}}{{\partial v}} - \frac{e}{{{m_e}}}\frac{{\vec B}}{c} \times {\vec f_1} = {\vec C_1}\left( {{{\vec f}_1}} \right),
\end{equation}
where ${C_0}\left( {{f_0}} \right)$ is replaced by the Bhatnagar-Gross-Kook (BGK) linear operator $ - {\nu _{ee}}\left( {{f_0} - f_0^m} \right)$, ${\vec C_1}\left( {{{\vec f}_1}} \right)$ can be simply approximated as $ - \nu _{ei}^*{\vec f_1}$, and $Z$ is the average ionization state. ${\nu _{ee}} = \frac{{4\pi {n_e}{e^4}\ln {\Lambda _{ee}}}}{{{v^3}m_e^2}}$, ${\nu _{ei}} = \frac{{4\pi {n_e}{e^4}Z\ln {\Lambda _{ei}}}}{{{v^3}m_e^2}}$ and $\nu _{ei}^* = \xi {\nu _{ei}} = \frac{{Z + 4.2}}{{Z + 0.24}}{\nu _{ei}}$ are the $e$-$e$ collision frequency, the $e$-$i$ collision frequency and the $e$-$i$ collision frequency applicable to arbitrary $Z$, respectively, where ${n_e}$ is the electron number density, with $\ln {\Lambda _{ee}}$ and $\ln {\Lambda _{ei}}$ the $e$-$e$ and $e$-$i$ Coulomb logarithms.

Assuming the electron distribution function is no longer Maxwellian but remains close to it, its deviation from equilibrium can be dealt with by perturbation theory
\begin{equation}\label{Eq5}
	\begin{split}
		&{f_0} = f_0^m + \Delta {f_0},\\
		&{{\vec f}_1} = \vec f_1^m + \Delta {{\vec f}_1},\\
		&\vec E = {{\vec E}^m} + \Delta \vec E,\\
		&\vec B = {{\vec B}_0} + \Delta \vec B,
	\end{split}
\end{equation}
where the superscript $m$ denotes the Maxwellian distribution, $f_0^m = 4\pi {n_e}{\left( {\frac{1}{{\pi v_{2T}^2}}} \right)^{1.5}}{e^{ - {{\left( {\frac{v}{{{v_{2T}}}}} \right)}^2}}}$ is the Maxwellian electron distribution function, ${v_{2T}} = \sqrt {2{k_B}{T_e}/{m_e}} $ is the electron thermal velocity, ${k_B}$ is the Boltzmann constant, ${T_e}$ is the electron temperature, $\vec f_1^m$ is derived from the Braginskii model, ${\vec E^m} = {\vec E_B}$ is the Braginskii electric field, ${\vec B_0}$ is the unperturbed magnetic field, with $\Delta \vec E$ and $\Delta \vec B$ the perturbations in the electric and magnetic fields. Substituting Eq. \eqref{Eq5} into Eqs. \eqref{Eq3} and \eqref{Eq4} yield expressions for $\Delta {f_0}$ and $\Delta {\vec f_1}$
\begin{equation}\label{Eq6}
	\begin{split}
		v\nabla  \cdot \left( {\vec f_1^m + \Delta {{\vec f}_1}} \right) - \frac{{e{{\vec E}^m}}}{{{m_e}{v^2}}}\frac{\partial }{{\partial v}}\left( {{v^2}\vec f_1^m} \right) - \frac{{e{{\vec E}^m}}}{{{m_e}{v^2}}}\frac{\partial }{{\partial v}}\left( {{v^2}\Delta {{\vec f}_1}} \right) - \frac{{e\Delta \vec E}}{{{m_e}{v^2}}}\frac{\partial }{{\partial v}}\left( {{v^2}\vec f_1^m} \right)\\
		- \frac{{e\Delta \vec E}}{{{m_e}{v^2}}}\frac{\partial }{{\partial v}}\left( {{v^2}\Delta {{\vec f}_1}} \right) =  - {\nu _{ee}}\Delta {f_0},
	\end{split}
\end{equation}
\begin{equation}\label{Eq7}
	\begin{split}
		\frac{v}{3}\nabla \Delta {f_0} - \frac{{e{{\vec E}^m}}}{{3{m_e}}}\frac{{\partial \Delta {f_0}}}{{\partial v}} - \frac{{e\Delta \vec E}}{{3{m_e}}}\frac{{\partial f_0^m}}{{\partial v}} - \frac{{e\Delta \vec E}}{{3{m_e}}}\frac{{\partial \Delta {f_0}}}{{\partial v}} - {\omega _0}\vec b \times \Delta {\vec f_1} - \Delta \omega \vec b \times \left( {\vec f_1^m + \Delta {{\vec f}_1}} \right)\\
		=  - \nu _{ei}^*\Delta {\vec f_1},
	\end{split}
\end{equation}
where ${\omega _0} = e\left| {{{\vec B}_0}} \right|/{m_e}c$ is the electron gyrofrequency and $\vec b = {\vec B_0}/\left| {{{\vec B}_0}} \right|$ is the unit vector in magnetic field direction. For the perturbative equations, Nicolaï et al. introduced further simplifications \cite{nicolaiPracticalNonlocalModel2006}. Firstly, in Eqs. \eqref{Eq6} and \eqref{Eq7}, the terms involving the electric field ${\vec E_B}$ are scaled by a factor of $1/{v^2}$. This scaling makes the electric field terms negligible at high energy of electrons compared to spatial derivative terms or collisional terms. This omission of the electric field terms in Eqs. \eqref{Eq6} and \eqref{Eq7} can be compensated for by adopting a reduced source term and implementing a mean free path with electric field correction, respectively. Secondly, distortions in the low-energy part of the distribution function are small, allowing the perturbations of the electric and magnetic fields to be temporarily neglected, i.e., $\Delta \vec E = \Delta \vec B = 0$. The perturbation equations reduce to
\begin{equation}\label{Eq8}
	v\nabla  \cdot \left( {\vec f_1^m + \Delta {{\vec f}_1}} \right) =  - {\nu _{ee}}\Delta {f_0},
\end{equation}
\begin{equation}\label{Eq9}
	\frac{v}{3}\nabla \Delta {f_0} - {\omega _0}\vec b \times \Delta {\vec f_1} =  - \nu _{ei}^*\Delta {\vec f_1}.
\end{equation}
When $\vec b \bot \Delta {\vec f_1}$, the vector relation $\vec b \times \left( {\vec b \times \Delta {{\vec f}_1}} \right) =  - \Delta {\vec f_1}$ holds. Substituting the result of $\vec b \times $Eq. \eqref{Eq9} into Eq. \eqref{Eq9}, we obtain
\begin{equation}\label{Eq10}
	\frac{{\lambda _{ei}^*}}{3}\vec b \times \nabla \Delta {f_0} + {\chi _v}\Delta {\vec f_1} =  - \vec b \times \Delta {\vec f_1},
\end{equation}
thus
\begin{equation}\label{Eq11}
	\Delta {\vec f_1} = \frac{{ - \lambda _{ei}^*}}{{3\left( {1 + \chi _v^2} \right)}}\left( {\nabla \Delta {f_0} + {\chi _v}\vec b \times \nabla \Delta {f_0}} \right),
\end{equation}
where $\lambda _{ei}^* = v/\nu _{ei}^*$ is the $e$-$i$ mean free path and ${\chi _v} = {\omega _0}/\nu _{ei}^*$ is the magnetization for electrons with velocity $v$. Substituting Eq. \eqref{Eq11} into \eqref{Eq8} yields an equation about $\Delta {f_0}$
\begin{equation}\label{Eq12}
	\frac{{\Delta {f_0}}}{{{\lambda _{ee}}}} - \nabla  \cdot \frac{{\lambda _{ei}^*}}{{3\left( {1 + \chi _v^2} \right)}}\left( {\nabla \Delta {f_0} + {\chi _v}\vec b \times \nabla \Delta {f_0}} \right) =  - \nabla  \cdot \vec f_1^m,
\end{equation}
where ${\lambda _{ee}} = v/\left( {r{\nu _{ee}}} \right) = Z{\lambda _{ei}}/r$ is the $e$-$e$ mean free path, $r$ is a dimensionless correction factor for the Krook frequency. Multiplying Eq. \eqref{Eq12} by ${m_e}{v^5}/2$ and integrating over the velocity domain bounded by ${v_{g \pm 1/2}}$, yields the multigroup nonlocal equations applicable to magnetohydrodynamic (MHD) codes
\begin{equation}\label{Eq13}
	\frac{{r{H_g}}}{{Z{\lambda _{ei,g}}}} - \nabla  \cdot \frac{{\lambda _{ei,g}^*}}{{3\left( {1 + \chi _{g}^2} \right)}}\left( {\nabla {H_g} + {\chi _{g}}\vec b \times \nabla {H_g}} \right) =  - \frac{{{m_e}}}{2}\nabla  \cdot \int_{{v_{g - 1/2}}}^{{v_{g + 1/2}}} {{v^5}\vec f_1^m} dv,
\end{equation}
where ${H_g} = \frac{{{m_e}}}{2}\int_{{v_{g - 1/2}}}^{{v_{g + 1/2}}} {{v^5}\Delta {f_0}dv} $ is the solution of the multigroup diffusion equation, ${\lambda _{ei,g}} = 4{\left( {\frac{{{e_g}}}{{{k_B}{T_e}}}} \right)^2}\frac{{{{\left( {{k_B}{T_e}} \right)}^2}}}{{4\pi Z{n_e}{e^4}\ln {\Lambda _{ei}}}}$ is the $e$-$i$ mean free path of the gth group of electrons, $\lambda _{ei,g}^* = {\lambda _{ei,g}}/\xi $ and ${\chi _{g}} = {\omega _0}\frac{{\lambda _{ei,g}^*}}{{{v_g}}} = 2\sqrt 2 {\left( {\frac{{{e_g}}}{{{k_B}{T_e}}}} \right)^{1.5}}\frac{{{\omega _0}\sqrt {{m_e}} {{\left( {{k_B}{T_e}} \right)}^{1.5}}}}{{\xi 4\pi Z{n_e}{e^4}\ln {\Lambda _{ei}}}}$. The total heat flux is corrected to
\begin{equation}\label{Eq14}
	\vec Q = \frac{{{m_e}}}{2}\int_0^\infty  {\left( {\vec f_1^m + \Delta {{\vec f}_1}} \right){v^5}dv}  = {\vec Q_B} - \sum\limits_g {\frac{{\lambda _{ei,g}^*}}{{3\left( {1 + \chi _{g}^2} \right)}}\left( {\nabla {H_g} + {\chi _{g}}\vec b \times \nabla {H_g}} \right)} ,
\end{equation}
where ${\vec Q_B}$ is the Braginskii heat flux,
\begin{equation}\label{Eq15}
	{\vec Q_B} =  - \left[ {{\kappa _\parallel }\vec b(\vec b \cdot \nabla {T_e}) + {\kappa _ \bot }\vec b \times (\nabla {T_e} \times \vec b) + {\kappa _ \wedge }(\vec b \times \nabla {T_e})} \right],
\end{equation}
${\kappa _\parallel }$ is the thermal conductivity parallel to the magnetic field, ${\kappa _ \bot }$ is the thermal conductivity perpendicular to the magnetic field but parallel to the temperature gradient, and ${\kappa _ \wedge }$ is the thermal conductivity perpendicular to both the magnetic field and the temperature gradient. 

In the preceding approximations, the influence of the electric field on the nonlocal heat conduction of electrons is neglected. A semiqualitative approach can partially compensate for the effect of electric fields. For Eq. \eqref{Eq4}, the term $\frac{{e\vec E}}{{3{m_e}}}\frac{{\partial {f_0}}}{{\partial v}}$ is replaced by $- \frac{{v\Delta \vec f_1^m}}{{{d_E}}}$ , where ${d_E} = \frac{{{m_e}{v^2}}}{{2e\left| {{{\vec E}_B}} \right|}}$ is the stopping length for electrons with energy ${e_g}$ in a constant electric field of magnitude $\left| {{{\vec E}_B}} \right|$ \cite{riquierMagneticFieldLaser2016}. Equations \eqref{Eq9}, \eqref{Eq10} and \eqref{Eq11} can be rewritten as 
\begin{equation}\label{Eq16}
	\frac{{\nabla \Delta {f_0}}}{3} - \frac{{{\chi _v}}}{{\lambda _{ei}^*}}\vec b \times \Delta {\vec f_1} =  - \frac{{\Delta {{\vec f}_1}}}{{\lambda _{ei}^E}},
\end{equation}
\begin{equation}\label{Eq17}
	\vec b \times \Delta {\vec f_1} =  - \lambda _{ei}^E\left[ {\frac{1}{3}\vec b \times \nabla \Delta {f_0} + \frac{{{\chi _v}}}{{\lambda _{ei}^*}}\Delta {{\vec f}_1}} \right],
\end{equation}
\begin{equation}\label{Eq18}
	\Delta {\vec f_1} = a_1^E\nabla \Delta {f_0} + a_2^E\vec b \times \nabla \Delta {f_0},
\end{equation}
where $\lambda _{ei}^E$ is the electron mean free path limited by electric field, $\frac{1}{{\lambda _{ei}^E}} = \frac{1}{{\lambda _{ei}^*}} + \frac{1}{{{d_E}}}$, ${a_1^E} = \frac{1}{{3\lambda _{ei}^E}}{\left[ {{{\left( {\frac{{{\chi _v}}}{{\lambda _{ei}^*}}} \right)}^2} + {{\left( {\frac{1}{{\lambda _{ei}^E}}} \right)}^2}} \right]^{ - 1}}$ and ${a_2^E} = \frac{{{\chi _v}}}{{3\lambda _{ei}^*}}{\left[ {{{\left( {\frac{{{\chi _v}}}{{\lambda _{ei}^*}}} \right)}^2} + {{\left( {\frac{1}{{\lambda _{ei}^E}}} \right)}^2}} \right]^{ - 1}}$. Therefore, the corresponding multigroup diffusion equations and heat flux formulas are expressed respectively as
\begin{equation}\label{Eq19}
	\frac{{r{H_g}}}{{Z{\lambda _{ei,g}}}} - \nabla  \cdot \left( {{a_{1,g}^E}\nabla {H_g} + {a_{2,g}^E}\vec b \times \nabla {H_g}} \right) =  - \frac{{{m_e}}}{2}\nabla  \cdot \int_{{v_{g - 1/2}}}^{{v_{g + 1/2}}} {{v^5}\vec f_1^m} dv,
\end{equation}
\begin{equation}\label{Eq20}
	\vec Q = \frac{{{m_e}}}{2}\int_0^\infty  {\left( {\vec f_1^m + \Delta {{\vec f}_1}} \right){v^5}dv}  = {\vec Q_B} - \sum\limits_g {\left( {{a_{1,g}^E}\nabla {H_g} + {a_{2,g}^E}\vec b \times \nabla {H_g}} \right)} ,
\end{equation}
where ${a_{1,g}^E} = \frac{1}{{3\lambda _{ei,g}^E}}{\left[ {{{\left( {\frac{{{\chi _{g}}}}{{\lambda _{ei,g}^*}}} \right)}^2} + {{\left( {\frac{1}{{\lambda _{ei,g}^E}}} \right)}^2}} \right]^{ - 1}}$, ${a_{2,g}^E} = \frac{{{\chi _{g}}}}{{3\lambda _{ei,g}^*}}{\left[ {{{\left( {\frac{{{\chi _{g}}}}{{\lambda _{ei,g}^*}}} \right)}^2} + {{\left( {\frac{1}{{\lambda _{ei,g}^E}}} \right)}^2}} \right]^{ - 1}}$ and $\frac{1}{{\lambda _{ei,g}^E}} = \frac{1}{{\lambda _{ei,g}^*}} + \frac{{e\left| {{{\vec E}_B}} \right|}}{{{e_g}}}$.

\subsection{Treatment of source term}
\label{subsec2.2}
According to Eq. \eqref{Eq4}, when the magnetic field is neglected, the anisotropic component of the distribution function $\vec f_1^m$ is expressed as
\begin{equation}\label{Eq21}
	\vec f_1^m =  - \frac{{\lambda _{ei}^*}}{3}\left( {\frac{{{m_e}{v^2}}}{{2{k_B}{T_e}}} - 4} \right)f_0^m\frac{{\nabla {T_e}}}{{{T_e}}}.
\end{equation}
To omit the electric field term of Eq. \eqref{Eq6}, Schurtz et al. disregarded the cold electron return current term $\frac{{{m_e}{v^2}}}{{2{k_B}{T_e}}} - 4$ in Eq. \eqref{Eq21}, yielding a simplified expression \cite{schurtzNonlocalElectronConduction2000,brodrickNonlocalTransportFusionRelevant2019}
\begin{equation}\label{Eq22}
	\vec g_1^m =  - \frac{{\lambda _{ei}^*}}{3}f_0^m\frac{{\nabla {T_e}}}{{{T_e}}}.
\end{equation}
Thus, the source term in the multigroup diffusion equations can be written as 
\begin{equation}\label{Eq23}
	 - \frac{{{m_e}}}{2}\nabla  \cdot \int_{{v_{g - 1/2}}}^{{v_{g + 1/2}}} {{v^5}\vec f_1^m} dv =  - \nabla  \cdot {\vec Q_{SH,g}} =  - \nabla  \cdot {\eta _{0,g}}{\vec Q_{SH}},
\end{equation}
\begin{equation}\label{Eq24}
	{\eta _{0,g}} = \frac{{\int_{{v_{g - 1/2}}}^{{v_{g + 1/2}}} {{v^5}\vec f_1^m} dv}}{{\int_0^\infty  {{v^5}\vec f_1^m} dv}} \approx \frac{{\int_{{v_{g - 1/2}}}^{{v_{g + 1/2}}} {{v^5}\vec g_1^m} dv}}{{\int_0^\infty  {{v^5}\vec g_1^m} dv}} = \frac{{\int_{{e_{g - 1/2}}/{k_B}{T_e}}^{{e_{g + 1/2}}/{k_B}{T_e}} {{\beta ^4}{e^{ - \beta }}d\beta } }}{{24}},
\end{equation}
where ${\vec Q_{SH}} = \frac{{{m_e}}}{2}\int_0^\infty  {{v^5}\vec f_1^m} dv$ is the Spitzer-H\"arm (SH) heat flux, $\beta  = \frac{{{e_g}}}{{{k_B}{T_e}}}$, and ${\eta _{0,g}}$ is the contribution weight of the $g$th group electrons to the total local heat flux. Additionally, the integrals $\int_0^\infty  {{v^5}\vec f_1^m} dv$ and $\int_0^\infty  {{v^5}\vec g_1^m} dv$ yield identical values, i.e., $-24$. 

For the magnetized plasmas, the anisotropic component $\vec f_1^m$ of the distribution function is
\begin{equation}\label{Eq25}
	\vec f_1^m =  - \frac{1}{{3\left( {\nu _{ei}^{*2} + \omega _0^2} \right)}}\left[ {\nu _{ei}^*v\nabla f_0^m - \nu _{ei}^*\frac{{e{{\vec E}^m}}}{{{m_e}}}\frac{{\partial f_0^m}}{{\partial v}} + v{\omega _0}\vec b \times \nabla f_0^m - {\omega _0}\vec b \times \frac{{e{{\vec E}^m}}}{{{m_e}}}\frac{{\partial f_0^m}}{{\partial v}}} \right].
\end{equation}
Nicolaï et al. suggest using the same source term as for the unmagnetized model. However, this simplification ignores the dependence of the source term on the magnetic field. For Eq. \eqref{Eq13}, the source term $ - \frac{{{m_e}}}{2}\nabla  \cdot \int_{{v_{g - 1/2}}}^{{v_{g + 1/2}}} {{v^5}\vec f_1^m} dv =  - \nabla  \cdot {\vec Q_{B,g}}$ corresponds to the divergence of the Braginskii heat flux for each electron group. The anisotropic heat flux ${\vec Q_{B,g}}$ consists of three directional components ${\vec Q_{\parallel ,g}}$, ${\vec Q_{ \bot ,g}}$ and ${\vec Q_{ \wedge ,g}}$. The magnitudes of ${\vec Q_{ \bot ,g}}$ and ${\vec Q_{ \wedge ,g}}$ are strongly dependent on the magnetic field. Substituting ${\vec Q_{SH,g}}$ for ${\vec Q_{B,g}}$ is equivalent to setting ${\vec Q_{ \wedge ,g}} = 0$, ${\vec E^m} \approx {\vec E_{SH}}$, ignoring the effects of the return current of the cold electrons and the magnetic field on ${\vec Q_{ \bot ,g}}$, i.e.,
\begin{equation}\label{Eq26}
	\vec g_1^m =  - \frac{1}{{3\left( {\nu _{ei}^{*2} + \cancel{{\omega _0^2}}} \right)}}\left[ {\nu _{ei}^*v\nabla f_0^m - \nu _{ei}^*\frac{{e{{\vec E}^m}}}{{{m_e}}}\frac{{\partial f_0^m}}{{\partial v}} + \cancel{{v{\omega _0}\vec b \times \nabla f_0^m}} - \cancel{{{\omega _0}\vec b \times \frac{{e{{\vec E}^m}}}{{{m_e}}}\frac{{\partial f_0^m}}{{\partial v}}}}} \right],
\end{equation}
\begin{equation}\label{Eq27}
	\nabla f_0^m - \frac{{e{{\vec E}^m}}}{{{m_e}v}}\frac{{\partial f_0^m}}{{\partial v}} \approx f_0^m\left[ {{{\left( {\frac{v}{{{v_{2T}}}}} \right)}^2} - 4} \right]\frac{{\nabla {T_e}}}{{{T_e}}} \approx f_0^m\frac{{\nabla {T_e}}}{{{T_e}}}.
\end{equation}
Through verification, this substitution leads to a larger source term, resulting in excessively nonlocal heat flux in Eqs. \eqref{Eq14} and \eqref{Eq20}, thereby inducing unphysical results. To address this, appropriate modifications to the model are required. While assuming ${\vec E^m} \approx {\vec E_{SH}}$ and ignoring the electron return current, we incorporated the Righi-Leduc flux ${\vec Q_{ \wedge ,g}}$ and the dependence of ${\vec Q_{ \bot ,g}}$ on the magnetic field. The revised expressions are
\begin{equation}\label{Eq28}
	\vec g_1^m =  - \frac{1}{3}\left[ {\frac{{\lambda _{ei}^*}}{{1 + \chi _v^2}}f_0^m\frac{{\nabla {T_e}}}{{{T_e}}} + \frac{{\lambda _{ei}^*{\chi _v}}}{{1 + \chi _v^2}}\vec b \times \left( {f_0^m\frac{{\nabla {T_e}}}{{{T_e}}}} \right)} \right],
\end{equation}
\begin{equation}\label{Eq29}
	- \frac{{{m_e}}}{2}\nabla  \cdot \int_{{v_{g - 1/2}}}^{{v_{g + 1/2}}} {{v^5}\vec g_1^m} dv =  - \nabla  \cdot \left( {{\eta _{1,g}}{{\vec Q}_{ \bot ,g}} + {\eta _{2,g}}{{\vec Q}_{ \wedge ,g}}} \right),
\end{equation}
\begin{equation}\label{Eq30}
	{\eta _{1,g}} \approx \frac{{\int_{{e_{g - 1/2}}/{k_B}{T_e}}^{{e_{g + 1/2}}/{k_B}{T_e}} {\frac{{{\beta ^4}{e^{ - \beta }}}}{{1 + \chi _v^2}}d\beta } }}{{24}},\ {\eta _{2,g}} \approx \frac{{{\omega _0}{{\left( {{k_B}{T_e}} \right)}^{1.5}}}}{{6\sqrt 2 {n_e}{m_e}Y_{ei}^*}}\int_{{e_{g - 1/2}}/{k_B}{T_e}}^{{e_{g + 1/2}}/{k_B}{T_e}} {\frac{{{\beta ^{11/2}}{e^{ - \beta }}}}{{1 + \chi _v^2}}d\beta } ,
\end{equation}
\begin{equation}\label{Eq31}
	{\vec Q_ \bot } = {\kappa _{SH}}\nabla {T_e},\ {\vec Q_ \wedge } = {\kappa _{SH}}\vec b \times \nabla {T_e},
\end{equation}
where $\lambda _{ei}^* \propto {\beta ^2}$, ${\chi _v} \propto {\beta ^{1.5}}$, and ${\kappa _{SH}}$ is the SH thermal conductivity. For ${\vec Q_{\parallel ,g}}$, the weighting coefficient remains as ${\eta _{0,g}}$. The modified source term clearly restores the dependence on the magnetic field and decreases dynamically as the magnetic field increases, resulting in a more accurate nonlocal heat flux.

\subsection{Nonlocal Biermann magnetic field}
\label{subsec2.3}
Initially, in the absence of a magnetic field, through Eq. \eqref{Eq4} and the zero current condition
\begin{equation}\label{Eq32}
	\vec J =  - e\int_0^\infty  {\vec f_1^m{v^3}dv}  = 0,
\end{equation}
the kinetic expression for the electric field that generates the Biermann magnetic field under the Maxwell distribution can be derived as
\begin{equation}\label{Eq33}
	\vec E_{Bier}^m =  - \frac{{{m_e}}}{{6e}}\frac{{\nabla \int_0^\infty  {f_0^m{v^7}dv} }}{{\int_0^\infty  {f_0^m{v^5}dv} }} =  - \frac{1}{e}\left( {\frac{{\nabla {P_e}}}{{{n_e}}} + \frac{3}{2}{k_B}\nabla {T_e}} \right),
\end{equation}
where the thermoelectric term $\frac{3}{2}{k_B}\nabla {T_e}$ is curl-free and is usually neglected. We introduce perturbations $\Delta {f_0}$, $\Delta {\vec f_1}$ and $\Delta \vec E$ into Eq. \eqref{Eq4} yields
\begin{equation}\label{Eq34}
	\frac{v}{3}\nabla \left( {f_0^m + \Delta {f_0}} \right) - \frac{{e\left( {{{\vec E}^m} + \Delta \vec E} \right)}}{{3{m_e}}}\frac{{\partial \left( {f_0^m + \Delta {f_0}} \right)}}{{\partial v}} =  - \nu _{ei}^*\left( {\vec f_1^m + \Delta {{\vec f}_1}} \right),
\end{equation}
while the zero current condition remains valid $\vec J = \Delta \vec J = 0$. The electric field is ultimately corrected to
\begin{equation}\label{Eq35}
	{\vec E_{Bier}} = \vec E_{Bier}^m + \Delta \vec E = \vec E_{Bier}^m\left[ {1 - \frac{{\int {\Delta {f_0}{v^5}dv} }}{{\int {\left( {f_0^m + \Delta {f_0}} \right){v^5}dv} }}} \right] - \frac{{{m_e}}}{{6e}}\frac{{\int {\nabla \Delta {f_0}{v^7}dv} }}{{\int {\left( {f_0^m + \Delta {f_0}} \right){v^5}dv} }}.
\end{equation}

The electric field becomes much more complex as a significant magnetic field is initially present. From Eq. (35) in Ref. \cite{nicolaiPracticalNonlocalModel2006}, the expression for the Biermann-producing electric field under a Maxwellian distribution in the initial presence of a magnetic field is 
\begin{equation}\label{Eq36}
	\frac{{e\vec E_{Bier}^m}}{{{m_e}}} = \frac{{\int_0^\infty  {\frac{{{v^6}}}{{1 + \chi _v^2}}\frac{{\partial f_0^m}}{{\partial v}}dv} \int_0^\infty  {\frac{{{v^7}\nabla f_0^m}}{{1 + \chi _v^2}}} dv + \int_0^\infty  {\frac{{{\chi _v}{v^6}}}{{1 + \chi _v^2}}\frac{{\partial f_0^m}}{{\partial v}}dv} \int_0^\infty  {\frac{{{\chi _v}{v^7}\nabla f_0^m}}{{1 + \chi _v^2}}} dv}}{{{{\left( {\int_0^\infty  {\frac{{{v^6}}}{{1 + \chi _v^2}}\frac{{\partial f_0^m}}{{\partial v}}dv} } \right)}^2} + {{\left( {\int_0^\infty  {\frac{{{\chi _v}{v^6}}}{{1 + \chi _v^2}}\frac{{\partial f_0^m}}{{\partial v}}dv} } \right)}^2}}}.
\end{equation}
Introducing the perturbation $\Delta {f_0}$ into Eq. \eqref{Eq36} yields
\begin{equation}\label{Eq37}
	\frac{{e{{\vec E}_{Bier}}}}{{{m_e}}} = \frac{{\int_0^\infty  {\frac{{{v^6}\frac{{\partial \left( {f_0^m + \Delta {f_0}} \right)}}{{\partial v}}}}{{1 + \chi _v^2}}dv} \int_0^\infty  {\frac{{{v^7}\nabla \left( {f_0^m + \Delta {f_0}} \right)}}{{1 + \chi _v^2}}} dv + \int_0^\infty  {\frac{{{\chi _v}{v^6}\frac{{\partial \left( {f_0^m + \Delta {f_0}} \right)}}{{\partial v}}}}{{1 + \chi _v^2}}dv} \int_0^\infty  {\frac{{{\chi _v}{v^7}\nabla \left( {f_0^m + \Delta {f_0}} \right)}}{{1 + \chi _v^2}}} dv}}{{{{\left[ {\int_0^\infty  {\frac{{{v^6}}}{{1 + \chi _v^2}}\frac{{\partial \left( {f_0^m + \Delta {f_0}} \right)}}{{\partial v}}dv} } \right]}^2} + {{\left[ {\int_0^\infty  {\frac{{{\chi _v}{v^6}}}{{1 + \chi _v^2}}\frac{{\partial \left( {f_0^m + \Delta {f_0}} \right)}}{{\partial v}}dv} } \right]}^2}}}.
\end{equation}
Next, we discuss specific expressions for the electric field. In the multigroup diffusion equations, ${H_g}$ is defined as ${H_g} = \frac{{{m_e}}}{2}\int_{{v_{g - 1/2}}}^{{v_{g + 1/2}}} {{v^5}\Delta {f_0}dv}$, from which we derive
\begin{equation}\label{Eq38}
	\begin{split}
		&\int_0^\infty  {\Delta {f_0}{v^5}dv}  \approx \frac{2}{{{m_e}}}\sum\limits_g {{H_g}} ,\\
		&\int_0^\infty  {\Delta {f_0}{v^7}dv}  \approx \frac{2}{{{m_e}}}\sum\limits_g {v_g^2{H_g}} .
	\end{split}
\end{equation}
Typically, the density perturbation $\Delta {n_e}$ caused by the distortion of the electron distribution function can be neglected
\begin{equation}\label{Eq39}
	\Delta {n_e} = \int_0^\infty  {\Delta {f_0}{v^2}dv}  \approx \frac{2}{{{m_e}}}\sum\limits_g {\frac{{{H_g}}}{{v_g^3}} \ll } {n_e}.
\end{equation}
However, for specific temperature-density configurations, density perturbations may cause the nonlocal Biermann magnetic field to deviate significantly from expected behavior. Assuming the total electron density is
\begin{equation}\label{Eq40}
	{n_e} = \int_0^\infty  {\left( {f_0^m + \Delta {f_0}} \right){v^2}dv} ,
\end{equation}
the electron distribution function should be modified to
\begin{equation}\label{Eq41}
	\begin{split}
		&f_0^m = 4\pi \left( {{n_e} - \Delta {n_e}} \right){\left( {\frac{1}{{\pi v_{2T}^2}}} \right)^{1.5}}{e^{ - {{\left( {\frac{v}{{{v_{2T}}}}} \right)}^2}}},\\
		&\int_0^\infty  {f_0^m{v^5}dv}  = \frac{{4v_T^3}}{{\sqrt \pi  }}\left( {{n_e} - \Delta {n_e}} \right).
	\end{split}
\end{equation}
The expression for the total electric field without the initial magnetic field is
\begin{equation}\label{Eq42}
	{\vec E_{Bier}} =  - \frac{{\nabla {P_e}}}{{e\left( {{n_e} - \Delta {n_e}} \right)}}\left[ {1 - \frac{{\sum\limits_g {{H_g}} }}{{\frac{{2{m_e}v_T^3}}{{\sqrt \pi  }}\left( {{n_e} - \Delta {n_e}} \right) + \sum\limits_g {{H_g}} }}} \right] - \frac{{{m_e}}}{{6e}}\frac{{\nabla \sum\limits_g {v_g^2{H_g}} }}{{\frac{{2{m_e}v_T^3}}{{\sqrt \pi  }}\left( {{n_e} - \Delta {n_e}} \right) + \sum\limits_g {{H_g}} }}.
\end{equation}
The magnetic field is then updated by Faraday’s law
\begin{equation}\label{Eq43}
	\frac{1}{c}\frac{{\partial \vec B}}{{\partial t}} =  - \nabla  \times \vec E.
\end{equation}

\subsection{Nonlocal Nernst velocity}
\label{subsec2.4}
The kinetic expression for the Nernst velocity \cite{lucianiMagneticFieldNonlocal1985,sherlockSuppressionBiermannBattery2020} is given by
\begin{equation}\label{Eq44}
	{\vec v_N} = \frac{{ - 3\int {{{\vec f}_1}{v^6}dv} }}{{\int {\frac{{\partial {f_0}}}{{\partial v}}{v^6}dv} }}.
\end{equation}
By introducing the perturbations $\Delta {f_0}$ and $\Delta {\vec f_1}$ of the distribution function into Eq. \eqref{Eq44}, we obtain 
\begin{equation}\label{Eq45}
	{\vec v_N} \approx \vec v_N^m + \frac{{\int {\frac{{ - \lambda _{ei}^*}}{{3\left( {1 + \chi _v^2} \right)}}\nabla \Delta {f_0}{v^6}dv} }}{{2\int {\left( {f_0^m + \Delta {f_0}} \right){v^5}dv} }} + \frac{{\int {\frac{{ - \lambda _{ei}^*}}{{3\left( {1 + \chi _v^2} \right)}}{\chi _v}\vec b \times \nabla \Delta {f_0}{v^6}dv} }}{{2\int {\left( {f_0^m + \Delta {f_0}} \right){v^5}dv} }},
\end{equation}
where $\Delta {\vec f_1}$ is provided by Eq. \eqref{Eq11}, $\vec v_N^m =  - \frac{{c{\beta _ \wedge }}}{{e\left| {\vec B} \right|}}{k_B}\nabla {T_e}$ is the classical Nernst velocity, the second and third terms represent the corrections due to nonlocal effects on the Nernst velocity and the cross-gradient-Nernst velocity, respectively, and ${\beta _ \wedge }$ denotes the transverse thermoelectric coefficient. Since ${\beta _ \wedge }\left( {Z,{\chi _e}} \right)$ correlates with both the average ionization state and magnetization, to accurately capture the dependence of the nonlocal Nernst velocity on these parameters, we implement the following simple modification
\begin{equation}\label{Eq46}
	{\vec v_N} \approx \vec v_N^m + \frac{{{\beta _ \wedge }\left( {Z,{\chi _e}} \right)}}{{{\beta _ \wedge }\left( { + \infty ,{\chi _e}} \right)}}\frac{{\int {\frac{{ - {\lambda _{ei}}}}{{3\left[ {1 + {{\left( {{\chi _v}\xi } \right)}^2}} \right]}}\nabla \Delta {f_0}{v^6}dv}  + \int {\frac{{ - {\lambda _{ei}}}}{{3\left[ {1 + {{\left( {{\chi _v}\xi } \right)}^2}} \right]}}{\chi _v}\xi \vec b \times \nabla \Delta {f_0}{v^6}dv} }}{{2\int {\left( {f_0^m + \Delta {f_0}} \right){v^5}dv} }}.
\end{equation}

\section{Numerical implementation}
\label{sec3}
We have implemented the improved mSNB model, incorporating the implicit iterative algorithm by Cao et al. \cite{caoImprovedNonlocalElectron2015}, into the FLASH code. FLASH is a publicly available, high-performance computing, multiphysics, adaptive mesh refinement, finite-volume Eulerian hydrodynamics, and magnetohydrodynamics (MHD) code \cite{fryxellFLASHAdaptiveMesh2000,dubeyExtensibleComponentbasedArchitecture2009,leeSolutionAccurateEfficient2013,tzeferacosFLASHMHDSimulations2015a}. The FLASH code has been extensively validated and is widely used in scientific research.

The electron anisotropic heat conduction equation for magnetized plasmas is rewritten to
\begin{equation}\label{Eq47}
	\rho {c_v}\frac{{T_e^{n + 1} - T_e^n}}{{\Delta t}} =  - \nabla  \cdot Q_B^k + \nabla  \cdot Q_B^{k - 1} + \frac{r}{Z}\sum\nolimits_g {\frac{{H_g^{k - 1}}}{{\lambda _{ei,g}^{k - 1}}}} ,
\end{equation}
where the superscript $n$ is the time index with $n + 1 = k$, the superscript $k$ is the iteration step, $\Delta t$ is the time step size, and $H_g$ is obtained by solving Eq. \eqref{Eq13} or \eqref{Eq19}. The convergence criterion for the implicit algorithm is $\left| {\nabla  \cdot Q_B^k - \nabla  \cdot Q_B^{k - 1}} \right| \le \alpha_0 \rho {c_v}T_e^k/\Delta t$, where $\alpha_0$ is the convergence tolerance factor (typically set to 0.01), $\rho$ is the fluid density, and $c_v$ is the electron specific heat. Upon satisfying the convergence criterion, the loop terminates to finalize the heat conduction equation computation. This study focuses on two-dimensional (2D) simulations in the $xy$-plane, where the magnetic field is typically aligned along the $z$-axis. Consequently, the parallel heat flux term ${\kappa _\parallel }\vec b(\vec b \cdot \nabla {T_e})$ in ${\vec Q_B}$ can be neglected. 

The FLASH code extends the induction equation
\begin{equation}\label{Eq48}
	\frac{{\partial \vec B}}{{\partial t}} + \nabla  \times \left( { - \vec u \times \vec B} \right) + \nabla  \times \left( { - \frac{{c\nabla {p_e}}}{{e{n_e}}} - {{\vec v}_N} \times \vec B} \right) = 0,
\end{equation}
and energy equation
\begin{equation}\label{Eq49}
	\frac{{\partial \varepsilon }}{{\partial t}} + \nabla  \cdot \left[ {\left( {\varepsilon  + {p_{tot}}} \right)\vec u - \left( {\vec u \cdot \vec B} \right)\vec B} \right] - \nabla  \cdot \left[ {\vec B \times \left( { - \frac{{c\nabla {p_e}}}{{e{n_e}}} - {{\vec v}_N} \times \vec B} \right)} \right] =  - \nabla  \cdot {\vec Q_{tot}} + S,
\end{equation}
to incorporate MHD non-ideal terms including the Biermann battery magnetic field and Nernst effect, where $\vec u$ is the fluid velocity, $\eta $ is the magnetic resistivity, $\varepsilon $ is the total energy density, ${p_{tot}}$ is the total pressure, ${\vec Q_{tot}}$ is the total heat flux including the Ettingshausen effect, and $S$ is an additional source term. Modifications of magnetic fields induced by the electron nonlocal transport were implemented within the FLASH framework. However, we note that applying the current discretization scheme to solve the anisotropic heat conduction equation directly in FLASH is numerically unstable. To address this limitation, we conducted stability analysis of the anisotropic heat conduction equation and developed a numerically stable discretization scheme.

\subsection{Discretization scheme for the anisotropic heat conduction equation}
\label{subsec3.1}
The FLASH code adopts a staggered mesh finite volume method to discretize the first-order time implicit heat conduction equation based on Cartesian coordinates
\begin{equation}\label{Eq50}
	\begin{split}
		&\left( {\rho {c_v}} \right)_{i,j}^n\frac{{T_{i,j}^{n + 1} - T_{i,j}^n}}{{\Delta t}}\Delta V\\
		&= \left[ {{{\left( {\kappa _ \bot ^n\frac{{\partial {T^{n + 1}}}}{{\partial x}} - \kappa _ \wedge ^nb_z^n\frac{{\partial {T^{n + 1}}}}{{\partial y}}} \right)}_{i + 1/2,j}} - {{\left( {\kappa _ \bot ^n\frac{{\partial {T^{n + 1}}}}{{\partial x}} - \kappa _ \wedge ^nb_z^n\frac{{\partial {T^{n + 1}}}}{{\partial y}}} \right)}_{i - 1/2,j}}} \right]\Delta y \Delta z\\
		&+ \left[ {{{\left( {\kappa _ \bot ^n\frac{{\partial {T^{n + 1}}}}{{\partial y}} + \kappa _ \wedge ^nb_z^n\frac{{\partial {T^{n + 1}}}}{{\partial x}}} \right)}_{i,j + 1/2}} - {{\left( {\kappa _ \bot ^n\frac{{\partial {T^{n + 1}}}}{{\partial y}} + \kappa _ \wedge ^nb_z^n\frac{{\partial {T^{n + 1}}}}{{\partial x}}} \right)}_{i,j - 1/2}}} \right]\Delta x \Delta z\\
		&+ S_{i,j}^n,
	\end{split}
\end{equation}
where the subscripts $i$ and $j$ are spatial indices along the $x$- and $y$-axes, $\Delta V = \Delta x\Delta y\Delta z$ is the volume element, and $\Delta x$, $\Delta y$ and $\Delta z$ (with $\Delta z = 1$ for 2D simulations) represent the grid sizes in Cartesian coordinates. Similarly, the space discretization scheme for the multigroup diffusion equations is
\begin{equation}\label{Eq51}
	\begin{split}
		&r\left( {\frac{{{H_g}}}{{Z{\lambda _{ei,g}}}}} \right)_{i,j}^n \Delta V\\
		& - \left[ {\left( {{a_{1,g}^E}\frac{{\partial {H_g}}}{{\partial x}} - {a_{2,g}^E}{b_z}\frac{{\partial {H_g}}}{{\partial y}}} \right)_{i + 1/2,j}^n - \left( {{a_{1,g}^E}\frac{{\partial {H_g}}}{{\partial x}} - {a_{2,g}^E}{b_z}\frac{{\partial {H_g}}}{{\partial y}}} \right)_{i - 1/2,j}^n} \right]\Delta y \Delta z\\
		& - \left[ {\left( {{a_{1,g}^E}\frac{{\partial {H_g}}}{{\partial y}} + {a_{2,g}^E}{b_z}\frac{{\partial {H_g}}}{{\partial x}}} \right)_{i,j + 1/2}^n - \left( {{a_{1,g}^E}\frac{{\partial {H_g}}}{{\partial y}} + {a_{2,g}^E}{b_z}\frac{{\partial {H_g}}}{{\partial x}}} \right)_{i,j - 1/2}^n} \right]\Delta x \Delta z\\
		&  =  - {\left( {\int {\nabla  \cdot \vec Q_{B,g}^ndV} } \right)_{i,j}},
	\end{split}
\end{equation}
where
\begin{equation}\label{Eq52}
	\begin{split}
		& - {\left( {\int {\nabla  \cdot \vec Q_{B,g}^ndV} } \right)_{i,j}}=\\
		& - \left[ {\left( {{\eta _{1,g}}{\kappa _{SH}}\frac{{\partial T}}{{\partial x}} - {\eta _{2,g}}{\kappa _{SH}}{b_z}\frac{{\partial T}}{{\partial y}}} \right)_{i + 1/2,j}^n - \left( {{\eta _{1,g}}{\kappa _{SH}}\frac{{\partial T}}{{\partial x}} - {\eta _{2,g}}{\kappa _{SH}}{b_z}\frac{{\partial T}}{{\partial y}}} \right)_{i - 1/2,j}^n} \right]\Delta y\Delta z\\
		& - \left[ {\left( {{\eta _{1,g}}{\kappa _{SH}}\frac{{\partial T}}{{\partial y}} + {\eta _{2,g}}{\kappa _{SH}}{b_z}\frac{{\partial T}}{{\partial x}}} \right)_{i,j + 1/2}^n - \left( {{\eta _{1,g}}{\kappa _{SH}}\frac{{\partial T}}{{\partial y}} + {\eta _{2,g}}{\kappa _{SH}}{b_z}\frac{{\partial T}}{{\partial x}}} \right)_{i,j - 1/2}^n} \right]\Delta x\Delta z.
	\end{split}
\end{equation}
For the perpendicular components in Eq. \eqref{Eq50}, such as ${\left( {\kappa _ \bot ^n\frac{{\partial {T^{n + 1}}}}{{\partial x}}} \right)_{i + 1/2,j}}$, it can be discretized simply as
\begin{equation}\label{Eq53}
	{\left( {\kappa _ \bot ^n\frac{{\partial {T^{n + 1}}}}{{\partial x}}} \right)_{i + 1/2,j}} = \frac{{{{\left( {\kappa _ \bot ^n} \right)}_{i + 1,j}} + {{\left( {\kappa _ \bot ^n} \right)}_{i,j}}}}{2}\frac{{T_{i + 1,j}^{n + 1} - T_{i,j}^{n + 1}}}{{\Delta x}}.
\end{equation}
For cross components, slope limiters are usually required to ensure heat flux is transported from the region of high temperature to low temperature. Thus define
\begin{equation}\label{Eq54}
	{\left( {\kappa _ \wedge ^nb_z^n\frac{{\partial {T^{n + 1}}}}{{\partial y}}} \right)_{i + 1/2,j}} = \frac{{{{\left( {\kappa _ \wedge ^n} \right)}_{i + 1,j}} + {{\left( {\kappa _ \wedge ^n} \right)}_{i,j}}}}{2}\frac{{{{\left( {b_z^n} \right)}_{i + 1,j}} + {{\left( {b_z^n} \right)}_{i,j}}}}{2}{\left( {\frac{{\partial {T^{n + 1}}}}{{\partial y}}} \right)_{i + 1/2,j}},
\end{equation}
where
\begin{equation}\label{Eq55}
	\begin{split}
		&\left( {\frac{{\partial T}}{{\partial y}}} \right)_{i + 1/2,j}^{n + 1}\\
		&= L\left\{ {L\left[ {\left( {\frac{{\partial T}}{{\partial y}}} \right)_{i,j + 1/2}^{n + 1},\left( {\frac{{\partial T}}{{\partial y}}} \right)_{i,j - 1/2}^{n + 1}} \right],L\left[ {\left( {\frac{{\partial T}}{{\partial y}}} \right)_{i + 1,j + 1/2}^{n + 1},\left( {\frac{{\partial T}}{{\partial y}}} \right)_{i + 1,j - 1/2}^{n + 1}} \right]} \right\},
	\end{split}
\end{equation}
$L$ is a slope limiter, with common types including minmod, van Leer, and MC limiters. The expansion of other terms in Eq. \eqref{Eq50} follows a similar approach to Eqs. \eqref{Eq53} and \eqref{Eq54}.

\subsection{Stability analysis}
\label{subsec3.2}
When the anisotropic heat conduction equation contains only parallel and perpendicular terms, slope limiters effectively eliminate unphysical heat flux generated by the parallel term. However, when the cross term is included in the equation instead of the parallel term, the effectiveness of slope limiters requires further evaluation \cite{pertPhysicalConstraintsNumerical1981,sharmaPreservingMonotonicityAnisotropic2007}. The slope limiter in Eq. \eqref{Eq55} is equivalent to applying different weights to gradients at four distinct positions
\begin{equation}\label{Eq56}
	\begin{split}
		\left( {\frac{{\partial T}}{{\partial y}}} \right)_{i + 1/2,j}^{n + 1} &= {a_1}\left( {\frac{{\partial T}}{{\partial y}}} \right)_{i,j + 1/2}^{n + 1} + {a_2}\left( {\frac{{\partial T}}{{\partial y}}} \right)_{i + 1,j + 1/2}^{n + 1}\\
		&+ {a_3}\left( {\frac{{\partial T}}{{\partial y}}} \right)_{i,j - 1/2}^{n + 1} + {a_4}\left( {\frac{{\partial T}}{{\partial y}}} \right)_{i + 1,j - 1/2}^{n + 1},
	\end{split}
\end{equation}
where the values of ${a_1}$, ${a_2}$, ${a_3}$ and ${a_4}$ depend on the type of slope limiters. The weight coefficients of ${\frac{{\partial {T^{n + 1}}}}{{\partial y}}_{i - 1/2,j}}$, ${\frac{{\partial {T^{n + 1}}}}{{\partial x}}_{i,j + 1/2}}$ and ${\frac{{\partial {T^{n + 1}}}}{{\partial x}}_{i,j - 1/2}}$ are denoted as ${b_m}$, ${c_m}$ and ${d_m}$ respectively, where $m=1,...,4$. For simplicity, we consider a uniform 2D Cartesian grid ($\Delta x = \Delta y$ and $\Delta z = 1$). Equation \eqref{Eq50} can then be rewritten as
\begin{equation}\label{Eq57}
	\begin{split}
		T_{i,j}^{n + 1} - T_{i,j}^n &= {\alpha _1}{K_5}\left( {T_{i + 1,j}^{n + 1} - T_{i,j}^{n + 1}} \right) - {\alpha _1}{K_6}\left( {T_{i,j}^{n + 1} - T_{i - 1,j}^{n + 1}} \right)\\
		&- {\alpha _2}{K_1}{\Delta _1} + {\alpha _2}{K_2}{\Delta _2}\\
		&+ {\alpha _2}{K_7}\left( {T_{i,j + 1}^{n + 1} - T_{i,j}^{n + 1}} \right) - {\alpha _2}{K_8}\left( {T_{i,j}^{n + 1} - T_{i,j - 1}^{n + 1}} \right)\\
		&+ {\alpha _1}{K_3}{\Delta _3} - {\alpha _1}{K_4}{\Delta _4},
	\end{split}
\end{equation}
where ${\alpha _1} = \Delta t/\Delta x$, ${\alpha _2} = \Delta t/\Delta y$,
\begin{equation}\label{Eq58}
	\begin{split}
		&{K_1} = \left( {\frac{{{\kappa _ \wedge }{b_z}}}{{\rho {c_v}}}} \right)_{i + 1/2,j}^n\frac{{\Delta y\Delta z}}{{\Delta V}},{K_2} = \left( {\frac{{{\kappa _ \wedge }{b_z}}}{{\rho {c_v}}}} \right)_{i - 1/2,j}^n\frac{{\Delta y\Delta z}}{{\Delta V}},\\
		&{K_3} = \left( {\frac{{{\kappa _ \wedge }{b_z}}}{{\rho {c_v}}}} \right)_{i,j + 1/2}^n\frac{{\Delta x\Delta z}}{{\Delta V}},{K_4} = \left( {\frac{{{\kappa _ \wedge }{b_z}}}{{\rho {c_v}}}} \right)_{i,j - 1/2}^n\frac{{\Delta x\Delta z}}{{\Delta V}},\\
		&{K_5} = \left( {\frac{{{\kappa _ \bot }}}{{\rho {c_v}}}} \right)_{i + 1/2,j}^n\frac{{\Delta y\Delta z}}{{\Delta V}},{K_6} = \left( {\frac{{{\kappa _ \bot }}}{{\rho {c_v}}}} \right)_{i - 1/2,j}^n\frac{{\Delta y\Delta z}}{{\Delta V}},\\
		&{K_7} = \left( {\frac{{{\kappa _ \bot }}}{{\rho {c_v}}}} \right)_{i,j + 1/2}^n\frac{{\Delta x\Delta z}}{{\Delta V}},{K_8} = \left( {\frac{{{\kappa _ \bot }}}{{\rho {c_v}}}} \right)_{i,j - 1/2}^n\frac{{\Delta x\Delta z}}{{\Delta V}},
	\end{split}
\end{equation}
and
\begin{equation}\label{Eq59}
	\begin{split}
		&{\Delta _1} = {a_1}\left( {T_{i,j + 1}^{n + 1} - T_{i,j}^{n + 1}} \right) + {a_2}\left( {T_{i + 1,j + 1}^{n + 1} - T_{i + 1,j}^{n + 1}} \right) + {a_3}\left( {T_{i,j}^{n + 1} - T_{i,j - 1}^{n + 1}} \right) + {a_4}\left( {T_{i + 1,j}^{n + 1} - T_{i + 1,j - 1}^{n + 1}} \right),\\
		&{\Delta _2} = {b_1}\left( {T_{i - 1,j + 1}^{n + 1} - T_{i - 1,j}^{n + 1}} \right) + {b_2}\left( {T_{i,j + 1}^{n + 1} - T_{i,j}^{n + 1}} \right) + {b_3}\left( {T_{i - 1,j}^{n + 1} - T_{i - 1,j - 1}^{n + 1}} \right) + {b_4}\left( {T_{i,j}^{n + 1} - T_{i,j - 1}^{n + 1}} \right),\\
		&{\Delta _3} = {c_1}\left( {T_{i + 1,j + 1}^{n + 1} - T_{i,j + 1}^{n + 1}} \right) + {c_2}\left( {T_{i + 1,j}^{n + 1} - T_{i,j}^{n + 1}} \right) + {c_3}\left( {T_{i,j + 1}^{n + 1} - T_{i - 1,j + 1}^{n + 1}} \right) + {c_4}\left( {T_{i,j}^{n + 1} - T_{i - 1,j}^{n + 1}} \right),\\
		&{\Delta _4} = {d_1}\left( {T_{i + 1,j}^{n + 1} - T_{i,j}^{n + 1}} \right) + {d_2}\left( {T_{i + 1,j - 1}^{n + 1} - T_{i,j - 1}^{n + 1}} \right) + {d_3}\left( {T_{i,j}^{n + 1} - T_{i - 1,j}^{n + 1}} \right) + {d_4}\left( {T_{i,j - 1}^{n + 1} - T_{i - 1,j - 1}^{n + 1}} \right).
	\end{split}
\end{equation}

Applying von Neumann analysis to Eq. \eqref{Eq57}, we get
\begin{equation}\label{Eq60}
	\hat T_{i,j}^{n + 1} = G\left( {i,j,n,k,\Delta x,\Delta y,\Delta t} \right)\hat T_{i,j}^n,
\end{equation}
where $\hat T_{i,j}^n = {e^{I\left( {ik\Delta x + jk\Delta y} \right)}}$, $I = \sqrt { - 1} $, $k$ is the wavenumber, and the squared modulus of the amplification factor $G$ is
\begin{equation}\label{Eq61}
	{\left| {G\left( {i,j,n,k,\Delta x,\Delta y,\Delta t} \right)} \right|^2} = \frac{1}{{1 + {\alpha ^2}{\mu _1} + \alpha {\mu _2}}},
\end{equation}
with
\begin{equation}\label{Eq62}
	\begin{split}
		&{\mu _1} = {\left[ {2{t_2}{{\sin }^2}\left( {\frac{\theta }{2}} \right) + 2{t_3}{{\sin }^2}\left( \theta  \right)} \right]^2} + {\sin ^2}\left( \theta  \right){\left[ {{t_4} - 2{t_5}\cos \left( \theta  \right)} \right]^2},\\
		&{\mu _2} =  - 4{\sin ^2}\frac{\theta }{2}\left( {{t_2} + 2{t_3} + 2{t_3}\cos \theta } \right),
	\end{split}
\end{equation}
$\theta  = k\Delta x = k\Delta y$, $\alpha  = {\alpha _1} = {\alpha _2}$ and 
\begin{equation}\label{Eq63}
	\begin{split}
		{t_1} &= {K_1}({a_1} - {a_3} + {a_4}) + {K_2}({b_1} - {b_2} + {b_4}) + {K_3}( - {c_2} - {c_3} + {c_4}) + {K_4}({d_1} - {d_2} - {d_3})\\
		&- {K_5} - {K_6} - {K_7} - {K_8},\\
		{t_2} &= {K_1}({a_1} - {a_2} - {a_3} + {a_4}) + {K_2}({b_1} - {b_2} - {b_3} + {b_4}) + {K_3}({c_1} - {c_2} - {c_3} + {c_4})\\
		&+ {K_4}({d_1} - {d_2} - {d_3} + {d_4}) - {K_5} - {K_6} - {K_7} - {K_8},\\
		{t_3} &= {a_2}{K_1} + {b_3}{K_2} - {c_1}{K_3} - {d_4}{K_4},\\
		{t_4} &= {K_1}( - {a_1} + {a_2} - {a_3} - {a_4}) + {K_2}({b_1} + {b_2} - {b_3} + {b_4}) + {K_3}( - {c_1} + {c_2} + {c_3} + {c_4})\\
		&+ {K_4}( - {d_1} - {d_2} - {d_3} + {d_4}) + {K_5} - {K_6} + {K_7} - {K_8},\\
		{t_5} &= {a_2}{K_1} - {b_3}{K_2} - {c_1}{K_3} + {d_4}{K_4}.
	\end{split}
\end{equation}
Equation \eqref{Eq57} represents an implicit variable-coefficient discretization scheme. To ensure the stability of this scheme, the amplification factor $G\left( {i,j,n,\theta ,\alpha } \right)$ must satisfy the stability condition
\begin{equation}\label{Eq64}
	\mathop {\max }\limits_{i,j,n,\theta ,\alpha } \left[ {\left| {G\left( {i,j,n,\theta ,\alpha } \right)} \right|} \right] \le 1.
\end{equation}
This condition is equivalent to requiring ${\mu _1} \ge 0$ and ${\mu _2} \ge 0$ to hold universally. While ${\mu _1} \ge 0$ is always satisfied, Eq. \eqref{Eq62} indicates that ${\mu _2} \ge 0$ requires the simultaneous fulfillment of ${t_2} < 0$ and ${t_3} < 0$. This imposes constraints on the application of slope limiters.

To simplify the analysis, we take the minmod slope limiter as an example. For the coefficients ${a_m}$, the stability conditions require
\begin{equation}\label{Eq65}
	\begin{split}
		&{a_2} = {a_3} = 0,\ {K_1} < 0\\
		&\left( {{a_1},{a_2},{a_3},{a_4}} \right) = \left( {0,0,1,0} \right).\ {K_1} \ge 0
	\end{split}
\end{equation}
When ${K_1} < 0$, the values of $a_1$ and $a_4$ are determined by comparing the gradients at the two pertinent locations and selecting the one with the smaller absolute value, using the slope limiter. Similarly, the constraints for coefficients ${b_m}$, ${c_m}$, and ${d_m}$ are derived as
\begin{equation}\label{Eq66}
	\begin{split}
		&{b_2} = {b_3} = 0,\ {K_2} < 0\\
		&\left( {{b_1},{b_2},{b_3},{b_4}} \right) = \left( {0,1,0,0} \right),\ {K_2} \ge 0\\
		&\left( {{c_1},{c_2},{c_3},{c_4}} \right) = \left( {0,0,0,1} \right),\ {K_3} < 0\\
		&{c_1} = {c_4} = 0,\ {K_3} \ge 0\\
		&\left( {{d_1},{d_2},{d_3},{d_4}} \right) = \left( {1,0,0,0} \right),\ {K_4} < 0\\
		&{d_1} = {d_4} = 0.\ {K_4} \ge 0
	\end{split}
\end{equation}
Other slope limiters may also be applied, provided they ensure ${t_2} < 0$ and ${t_3} < 0$.

\section{Numerical tests}
\label{sec4}
\subsection{Quasi-two-dimensional temperature ramp relaxation}
\label{subsec4.1}
We first analyze the quasi-2D temperature ramp relaxation under an initial uniform magnetic field, adopting the same parameter settings as Brodrick et al. \cite{brodrickIncorporatingKineticEffects2018}. Fully ionized helium plasma ($Z=2$) and zirconium plasma ($Z=40$) were studied, with the electron density fixed at $5 \times {10^{20}}\ \rm{cm^{-3}}$ and the Coulomb logarithm fixed at $7.09$. The initial temperature profile, varying from $1\ \rm{keV}$ to $150\ \rm{eV}$, is defined by
\begin{equation}\label{Eq67}
	{T_e}\left( {{\rm{eV}}} \right) = 575 - 425\tanh \left( {x/{L_0}} \right),
\end{equation}
where the scale lengths $L_0$ for the helium and zirconium simulations are $50\ \rm{\mu m}$ and $17.3\ \rm{\mu m}$, respectively. The temperature profile is along the $x$-axis, while the magnetic field is aligned parallel to the $z$-axis. However, to conveniently capture the $y$-direction physical quantities induced by the magnetic field, the simulation is extended to the $xy$-plane, hence termed the quasi-2D simulation. The simulation domains in the $x$- and $y$-directions are $ \pm 7{L_0}$ and $ \pm {L_0}$, with reflective and periodic boundary conditions applied at the x and y boundaries, respectively. The electron velocity ${v_g}$ in the nonlocal model is uniformly divided into $15$ groups, ranging from $\sqrt {0.025 \times 2/{m_e}} $ to $\sqrt {20{k_B}{T_{\max }} \times 2/{m_e}} $. The corresponding electron energy group is ${e_g} = {m_e}v_g^2/2$. In the simulations, the dimensionless parameters $r = 2\xi $ or $r = 3\xi $ are adopted, and both the electron mean free path $\lambda _{ei,g}^E$ limited by the electric field and the nonlocal Nernst velocity \eqref{Eq46} are applied.

\begin{figure}[htbp]\centering
	\includegraphics[width=10cm]{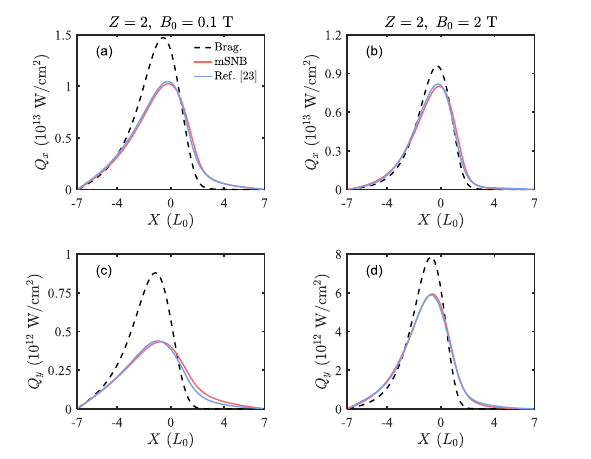}
	\caption{The perpendicular heat flux [(a) and (b)] and Righi-Leduc heat flux [(c) and (d)] in a helium plasma under magnetic fields of 0.1 T [(a) and (c), at 15 ps] and 2 T [(b) and (d), at 12 ps]. The black, red, and blue curves represent the results from the Braginskii model, the mSNB model ($r = 2\xi $), and the VFP model (K2 code; Brodrick et al. \cite{brodrickIncorporatingKineticEffects2018}), respectively.}\label{fig1}
\end{figure}

\begin{figure}[htbp]\centering
	\includegraphics[width=10cm]{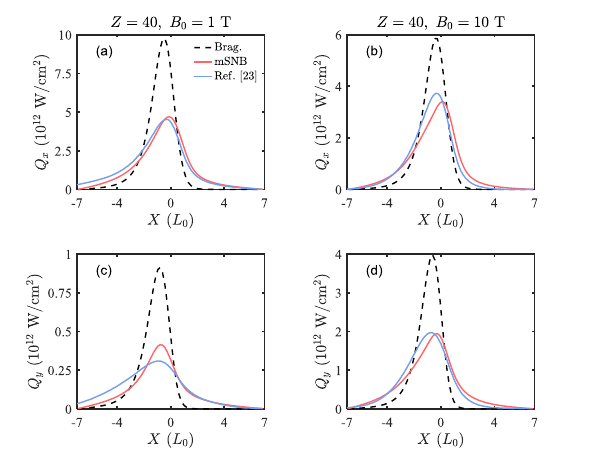}
	\caption{The perpendicular heat flux [(a) and (b)] and Righi-Leduc heat flux [(c) and (d)] in a zirconium plasma under magnetic fields of 1 T [(a) and (c), at 4 ps] and 10 T [(b) and (d), at 4 ps]. The black, red, and blue curves represent the results from the Braginskii model, the mSNB model ($r = 3\xi $), and the VFP model (K2 code; Brodrick et al. \cite{brodrickIncorporatingKineticEffects2018}), respectively.}\label{fig2}
\end{figure}

The simulation results shown in Figs. \ref{fig1} and \ref{fig2} demonstrate qualitatively consistent impacts of nonlocal electron heat conduction across varying ionization states. Nonlocal electron heat conduction reduces the peak heat flux of ${Q_x}$ and ${Q_y}$ while generating significant preheating at $2{L_0}$ to $4{L_0}$. The suppression effect of nonlocal conduction is more pronounced on the Righi-Leduc heat flux ($Q_y$). As the magnetic field strength increases, the nonlocality of electrons weakens, evidenced by the diminishing gap in peak heat flux of ${Q_x}$ and ${Q_y}$ between the mSNB model and the local Braginskii model. Our model demonstrates good agreement with the VFP model presented by Brodrick et al., validating the accuracy of the mSNB model. At $Z = 40$ and $B_z = 1\ \rm{T}$, the $Q_y$ heat flux in the mSNB model is slightly greater than that in the VFP model, probably due to the complex dependence of ${\kappa _ \wedge }$ on $Z$ and magnetization.

\begin{figure}[htbp]\centering
	\includegraphics[width=10cm]{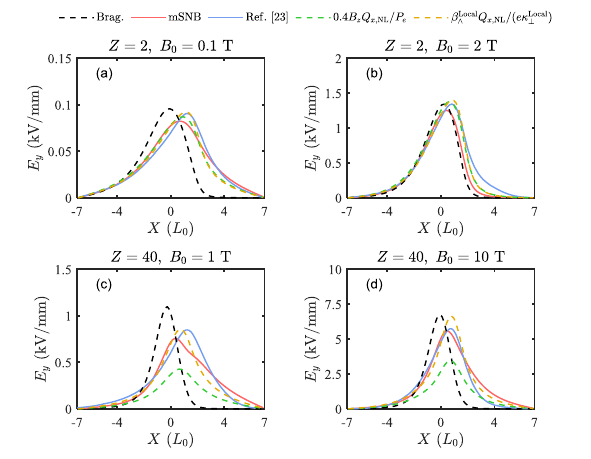}
	\caption{Electric field distributions under varying ionization states and magnetic fields. (a) $Z = 2$, ${B_0} = 0.1{\rm{\ T}}$. (b) $Z = 2$, ${B_0} = 2{\rm{\ T}}$. (c) $Z = 40$, ${B_0} = 1{\rm{\ T}}$. (d) $Z = 40$, ${B_0} = 10{\rm{\ T}}$. The green and yellow curves represent electric fields derived from two distinct approximations of the nonlocal Nernst velocity. The other curves are consistent with the description in Fig. \ref{fig1}.}\label{fig3}
\end{figure}

Figure \ref{fig3} shows the electric fields generated by Nernst advection under four conditions. Our results are consistent with the VFP model. Compared to the local electric field, the peak of the nonlocal electric field does not decrease significantly and shifts toward the cold region. This phenomenon is consistent with the conclusion of Brodrick et al. Adopting the nonlocal Nernst velocity ${\vec v_N} = 0.4{\vec Q_{x,NL}}/{P_e}$ is another common approximation method \cite{lanciaTopologyMegagaussMagnetic2014,hainesHeatFluxEffects1986}, which is reasonably effective for low-$Z$ materials but introduces significant underestimation at higher ionization levels. We also present the approximation ${\vec v_N} = \frac{{\beta _ \wedge ^{{\rm{Local}}}}}{{eB\kappa _ \bot ^{{\rm{Local}}}}}{\vec Q_{x,NL}}$ proposed by Brodrick et al., finding it likewise effective.

\subsection{Nonlocal suppression of the Biermann magnetic field}
\label{subsec4.2}
Under nonlocal regimes, less magnetic field generation is observed in experiments and VFP simulations than in classical MHD simulations \cite{sherlockSuppressionBiermannBattery2020,campbellMeasuringMagneticFlux2022,griff-mcmahonMeasurementsExtendedMagnetic2024a}. Davies established a fitting formula for the nonlocal suppression of the Biermann magnetic field valid for arbitrary atomic numbers using the 2D Fokker-Planck code K2 \cite{daviesNonlocalSuppressionBiermann2023}, distinguishing between cooling and heating scenarios. We adopted similar plasma density and temperature conditions
\begin{equation}\label{Eq68}
	{n_{e,0}} = 5 \times {10^{21}}\left[ {1 + {A_n}\cos \left( {{k_n}x} \right)} \right]\ {\rm{cm^{-3}}},
\end{equation}
\begin{equation}\label{Eq69}
	{T_{e,0}} = 2\left[ {1 + {A_T}\cos \left( {{k_T}y} \right)} \right]{\rm{\ keV,}}
\end{equation}
where ${k_T} = 1/{L_T}$, ${k_n} = {k_T}/512$, ${L_T}$ denotes the temperature gradient scale length, and $A_n$ and $A_T$ are the amplitudes of density and temperature perturbations, respectively. Reflective boundary conditions were applied along the $x$- and $y$-directions, with the simulation box dimensions set to half a period, i.e., $\pi {L_T} \times 512\pi {L_T}$. For the cooling case, we configured the initial density ${n_{e,0}}$ and temperature $T_{e,0}$ profiles with amplitudes ${A_n} = 0.1$ and ${A_T} = 0.5$. In the cooling mode, the nonlocal suppression factor $f_B^c$ for the Biermann magnetic field is defined as 
\begin{equation}\label{Eq70}
	f_B^c = \frac{{\max \left( {\left| {{B_{nonlocal}}} \right|} \right)}}{{\max \left( {\left| {{B_{class}}} \right|} \right)}}.
\end{equation}
For the heating case, the initial density profile ${n_{e,0}}$ was held constant, while the initial temperature was set to $2\left( {1 - {A_T}} \right){\rm{\ keV}}$ with amplitudes ${A_n} = 0.1$ and ${A_T} = 0.3$. The heating source was implemented as
\begin{equation}\label{Eq71}
	\frac{{\partial {T_e}}}{{\partial t}} = \frac{{{T_{e,0}} - {T_e}}}{{{\tau _h}}},
\end{equation}
where ${\tau _h} = 0.3{\rm{\ ps}}$ denotes the characteristic heating timescale. The nonlocal suppression factor $f_B^h$ for the Biermann magnetic field under the heating mode is defined as
\begin{equation}\label{Eq72}
	f_B^h = \frac{{\max \left[ {\left| {{{\left( {\partial B/\partial t} \right)}_{nonlocal}}} \right|} \right]}}{{\max \left[ {\left| {{{\left( {\partial B/\partial t} \right)}_{class}}} \right|} \right]}}.
\end{equation}
As electron heat conduction progresses, the temperature profile evolves toward spatial uniformity in the cooling case and stabilizes to a time-independent profile in the heating case. Consequently, ${f_B}$ gradually approaches a stable value in both cases, with this stabilized quantity adopted as the final result.

\begin{figure}[htbp]\centering
	\includegraphics[width=8cm]{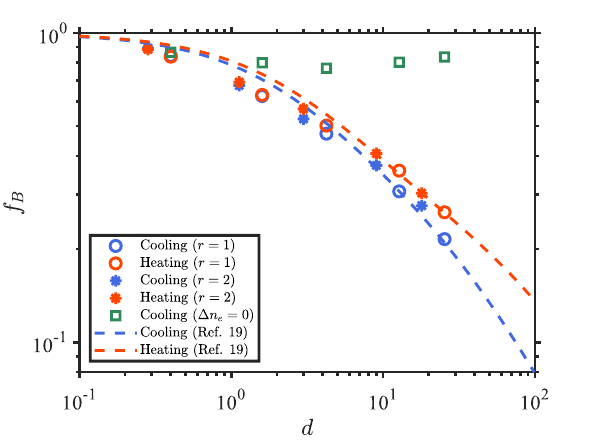}
	\caption{Suppression factor $f_B$ at $Z=1$ versus nonlocal parameter $d$, where open circles and asterisks denote results for $r=1$ and $r=2$, respectively, green squares represent cooling mode results with density perturbations ignored, the dashed curve corresponds to the fitting curve in Ref. \cite{daviesNonlocalSuppressionBiermann2023}, with blue and red colors indicating cooling and heating results.}\label{fig4}
\end{figure}

Simulations were performed for $Z = 1$ with temperature gradient scale lengths ${L_T} = 2,3.98,12,31.8$ and $127\ \rm{\mu m}$, retaining exclusively the Biermann magnetic field source term [implemented via Eq. \eqref{Eq35}] while disabling all other magnetic field generation modules. The results are shown in Fig. \ref{fig4}, where the dimensionless numbers $r = 1,2$ and the nonlocal parameter $d = {k_T}{l_d}$ is defined by ${l_d} = \sqrt {\frac{Z}{{\xi r}}} {l_{ei}}$ and ${l_{ei}} = \frac{8}{\pi }\sqrt {\frac{2}{\pi }} \frac{{{{\left( {{k_B}{T_e}} \right)}^2}}}{{Z{e^4}{n_e}\ln {\Lambda _{ei}}}}$. When the nonlocality is weak, the suppression factor $f_B$ approaches unity. Our results are marginally below the fitted curve in Ref. \cite{daviesNonlocalSuppressionBiermann2023}. As the nonlocality intensifies, its suppressive effect on the Biermann magnetic field strengthens, and our model demonstrates excellent agreement with the fitting curve. When density perturbations $\Delta {n_e}$ are neglected in the electric field equation \eqref{Eq42}, the suppression factor $f_B$ initially decreases with increasing $d$, then increases, yet remains above $0.75$. This result shows a significant deviation from the fitted curve. Overall, our model demonstrates reasonable agreement with the fitting curve given by Davies, lending further support to the validity of the mSNB model.

\subsection{Simplified laser-solid simulations}
\label{subsec4.3}
The Biermann battery mechanism generates a toroidal magnetic field around the laser spot during laser-solid interaction. The Nernst effect advects this field down the temperature gradients, causing it to propagate radially away from the laser spot. To investigate the impact of nonlocal electron transport on magnetic fields generated by laser-solid interactions, simulations were performed using parameters similar to those employed by Epperlein et al. in their benchmark tests \cite{epperleinTwoDimensionalNonlocalElectron1988,riquierMagneticFieldLaser2016}. The background plasma is fully ionized beryllium ($Z = 4$) with an initial temperature of $250\ \rm{eV}$. The electron density drops exponentially from $10^{23}\ \rm{cm^{-3}}$ at $x = - 25\ \rm{\mu m}$ to $4.2 \times 10^{20}\ \rm{cm^{-3}}$ at $x = - 150\ \rm{\mu m}$. Outside these endpoints ($x<- 150\ \rm{\mu m}$ and $x > - 25\ \rm{\mu m}$), the electron density is held constant at the respective endpoint values. A $3\omega_L$ laser pulse ($\lambda_L=351\ \rm{nm}$) with an intensity of ${I_0} = 5 \times {10^{14}}\ \rm{W/cm^2}$ irradiates the target from the left. The spatial profile of the laser intensity is given by $I = {I_0}\left[ {1 + \cos \left( {2\pi y/150\ \rm{\mu m}} \right)} \right]$, with the laser spot diameter of $150\ \rm{\mu m}$. The laser intensity is held constant in time. The simulation employed a nonlocal dimensionless parameter $r = 2$, incorporated electric field correction for the mean free path, applied nonlocal Biermann battery magnetic field [Eq. \eqref{Eq35}] and nonlocal Nernst effect, with a grid resolution of $2\ \rm{\mu m}$.

\begin{figure}[h]\centering
	\includegraphics[width=10cm]{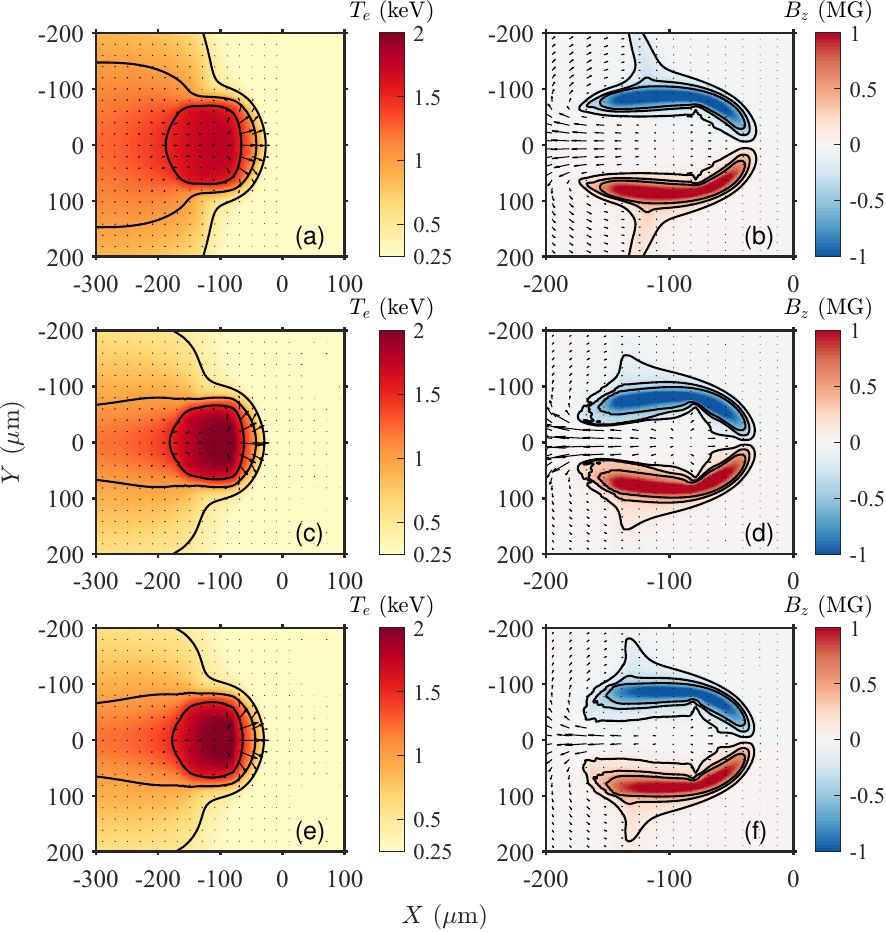}
	\caption{Temperature and magnetic field distributions at $100\ \rm{ps}$. (a) and (b) the Braginskii model. (c) and (d) the model with nonlocal thermal conduction and classical magnetic fields. (e) and (f) the model with nonlocal thermal conduction and nonlocal magnetic fields. Temperature contours correspond to $0.5$, $1$, and $1.5\ \rm{keV}$. Magnetic field contours denote $0.1$, $0.3$, and $0.5\ \rm{MG}$. Black arrows represent uniformly scaled heat flux in left panels and Nernst velocity in right panels.}\label{fig5}
\end{figure}

The results of the Braginskii model are shown in Figs. \ref{fig5}(a) and \ref{fig5}(b). Its magnetic field is more concentrated on both sides of the laser spot, effectively suppressing perpendicular heat conduction in these regions. This concentration confines energy within the focal spot radius, preventing coronal heating at larger radial distances along the $y$-direction. The Braginskii simulation omits flux limiters, leading to an overestimation of the coronal heat flux. This results in accelerated thermal diffusion from the corona toward overdense regions, ultimately yielding a lower peak temperature (approximately $1.8\ \rm{keV}$).

Results incorporating nonlocal thermal conduction and classical magnetic fields (including the Biermann battery and Nernst effect) are presented in Figs. \ref{fig5}(c) and \ref{fig5}(d). Owing to heat flux limitations, the coronal temperature now peaks at a higher value of $2.1\ \rm{keV}$. While nonlocal effects suppress heat flux in the corona region, the total heat flux marginally exceeds that in the Braginskii model because the local heat flux here is greater than in the Braginskii case. A similar pattern exists for the Nernst velocity. Compared to the Braginskii model, the magnetic field now exhibits a more diffuse distribution.

Figs. \ref{fig5}(e) and \ref{fig5}(f) present the results obtained with the combined implementation of nonlocal thermal conduction and nonlocal magnetic fields. As shown in Figs. \ref{fig5}(d) and \ref{fig5}(f), the radial distribution of nonlocal magnetic fields extends over a broader range while further suppressing the Nernst velocity in the corona. The heat flux in the corona undergoes modification due to this magnetic field advection, whereas the temperature profile remains less affected by nonlocal magnetic effects. Consequently, the coronal peak temperature maintains $2.1\ \rm{keV}$. Notably, nonlocal magnetic effects induce magnetic suppression, yielding peak magnetic field strengths of $1.33\ \rm{MG}$, $1.39\ \rm{MG}$, and $1.3\ \rm{MG}$ for the three respective cases. 

\begin{figure}[h]\centering
	\includegraphics[width=10cm]{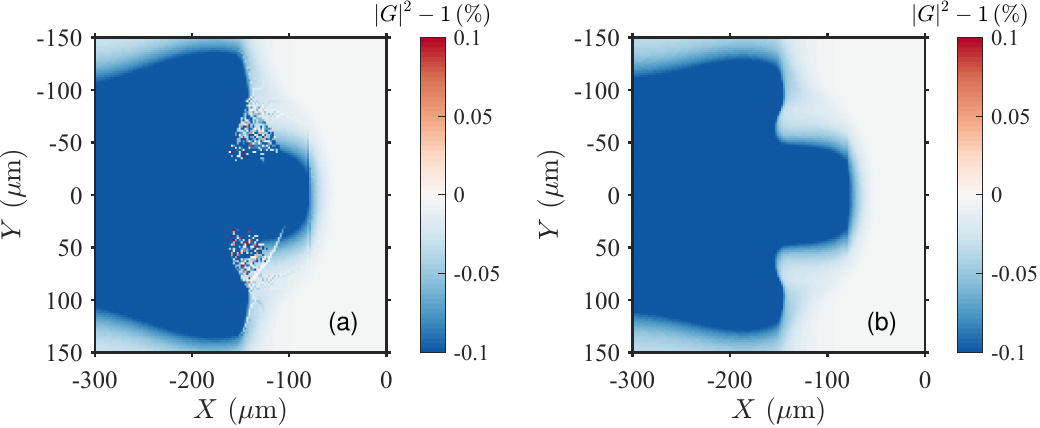}
	\caption{Squared magnitude of the amplification factor $G$ in the Braginskii model at $25\ \rm{ps}$ with $\theta  = 0.01$. (a) With a conventional minmod slope limiter. (b) With a stability-constrained minmod slope limiter.}\label{fig6}
\end{figure}

Figure \ref{fig6} compares results from the Braginskii model at $25\ \rm{ps}$ for the cases without (Fig. \ref{fig6}a) and with (Fig. \ref{fig6}b) the stability-constrained slope limiter. In Fig. \ref{fig6}(a), we employ the conventional minmod slope limiter, and some grid points violating the stability condition (i.e., ${\left| G \right|^2} > 1$) appear on both sides of the laser spot. This induces an unphysical heat flux that flows from colder to hotter regions, causing extreme temperatures that terminate the simulation. Conversely, the stability-constrained minmod slope limiter in Fig. \ref{fig6}(b) ensures ${\left| G \right|^2} < 1$ throughout the domain. Note that all results in Fig. \ref{fig5} were generated using this stability-constrained minmod limiter.

\section{Discussion}
\label{sec5}
The above simulations demonstrate that the improved multigroup nonlocal model for magnetized plasmas, derived from simplified VFP equations, can accurately capture electron nonlocal transport characteristics in magnetized plasmas.

For nonlocal thermal conduction under magnetic fields, our results establish that the Krook frequency correction factor $r = 2/\xi $ demonstrates better performance for low-$Z$ plasmas ($Z = 1$), while $r = 3/\xi $ yields a relatively improved accuracy of the model for mid-$Z$ plasmas ($Z = 40$). The model exhibits higher predictive fidelity for perpendicular heat flux than for Righi–Leduc heat flux, though its precision moderately diminishes with increasing the average ionization state $Z$. As the magnetic field increases, the reduction of plasma nonlocality leads to improvements in the model accuracy. Optimal selection of the correction factor $r$ for specific-$Z$ plasmas requires further investigation.

Regarding the nonlocal Nernst velocity, the model correctly captures its nonlocal characteristic that the electric field intensity peak position shifts toward the colder region. As magnetic fields strengthen, reduced nonlocality allows the model to recover classical Nernst velocity profiles.

In the case of the nonlocal Biermann battery, when density perturbations from nonlocality are neglected, increasing the parameter $d$ causes the magnetic field to dip to a minimum before recovering at larger $d$. The magnetic suppression by nonlocal effects does not intensify as expected with progressive enhancement of nonlocality, underscoring the critical role of density perturbations. Results for different correction factors $r$ lie on a single curve, demonstrating that the nonlocality must be characterized in conjunction with $r$ in this model. It should be noted that in other test cases, density perturbations are typically sufficiently small to be neglected.

In the simplified laser-solid interaction simulation, the model effectively captures the combined effects of nonlocal electron transport, including contributions from both the heat flux and magnetic fields. Theoretically, any physical process dependent on the temperature gradient is influenced by the nonlocal electron transport.

However, this model still exhibits certain limitations. Primarily, the neglect of the classical electric field ${\vec E_B}$ in Eq. \eqref{Eq7} could introduce inaccuracies. If we reintroduce ${\vec E_B}$ while disregarding perturbations in electric and magnetic fields ($\Delta \vec E$ and $\Delta \omega $), Eq. \eqref{Eq7} reduces to
\begin{equation}\label{Eq73}
	\frac{v}{3}\nabla \Delta {f_0} - \frac{{e{{\vec E}^m}}}{{3{m_e}}}\frac{{\partial \Delta {f_0}}}{{\partial v}} - {\omega _0}\vec b \times \Delta {\vec f_1} =  - \nu _{ei}^*\Delta {\vec f_1}.
\end{equation}
Applying the same methodology as in Section \ref{subsec2.1}, we derive an equation for $\Delta {f_0}$ and the corresponding multigroup equation
\begin{equation}\label{Eq74}
	\begin{split}
		&\frac{{\Delta {f_0}}}{{{\lambda _{ee}}}} - \nabla  \cdot \frac{{\lambda _{ei}^*}}{{3\left( {1 + \chi _v^2} \right)}}\left( {\nabla \Delta {f_0} + {\chi _v}\vec b \times \nabla \Delta {f_0}} \right) + \nabla  \cdot \frac{{\lambda _{ei}^*}}{{3\left( {1 + \chi _v^2} \right)}}\frac{e}{{{m_e}v}}\frac{{\partial \Delta {f_0}}}{{\partial v}}\left( {{{\vec E}^m} + {\chi _v}\vec b \times {{\vec E}^m}} \right)\\
		&=  - \nabla  \cdot \vec f_1^m,
	\end{split}
\end{equation}
\begin{equation}\label{Eq75}
	\begin{split}
		&\frac{{r{H_g}}}{{Z{\lambda _{ei,g}}}} - \nabla  \cdot \left[ {\frac{{\lambda _{ei,g}^*\left( {\nabla  + {\chi _g}\vec b \times \nabla } \right)}}{{3\left( {1 + \chi _g^2} \right)}}{H_g}} \right]\\
		&+ \nabla  \cdot \left[ {\frac{{\lambda _{ei,g}^*}}{{3\left( {1 + \chi _g^2} \right)}}\frac{{e\left( {{{\vec E}^m} + {\chi _g}\vec b \times {{\vec E}^m}} \right)}}{{{m_e}}}\left( {\frac{{{H_{g + 1}}}}{{{v_{g + 1}}}} - \frac{{{H_{g - 1}}}}{{{v_{g - 1}}}} - \frac{{4{H_g}}}{{v_g^2}}} \right)} \right] =  - \frac{{{m_e}}}{2}\nabla  \cdot \int_{{v_{g - 1/2}}}^{{v_{g + 1/2}}} {{v^5}\vec f_1^m} dv.
	\end{split}
\end{equation}
Equation \eqref{Eq75} self-consistently incorporates electric field effects without resorting to the semiqualitative correction \eqref{Eq19}, representing a key objective for our subsequent work.

In future work, other magnetic field terms, such as the cross-gradient-Nernst velocity and magnetic diffusion, will be investigated. Although current research on the impact of electron nonlocal transport on these two effects remains limited, and they are predicted to be less susceptible to nonlocality, further studies could be conducted to ascertain the extent of nonlocal effects on them.

\section{Conclusion}
\label{sec6}
The nonlocal multigroup model incorporating magnetic fields has been improved to accurately describe the electron nonlocal transport phenomena in magnetized plasmas. Firstly, the source term of the multigroup diffusion equation was modified to restore its dependence on the magnetic field, ensuring the model correctly calculates heat flux under varying magnetization strengths. Secondly, a nonlocal correction suitable for the multigroup diffusion model was proposed for the Biermann battery, adding nonlocal electric fields and density perturbations to the Biermann-producing electric field equation. Additionally, we introduced nonlocal Nernst velocity corrections that incorporate both the nonlocal Nernst velocity and the nonlocal cross-gradient-Nernst velocity. We also modified the numerical implementation methods. Firstly, the convergence criterion for the implicit iterative method of the multigroup model was improved to relate to the Braginskii heat flux. Then, we analyzed the stability of the discretization scheme for the anisotropic heat conduction equation. Specifically, the Righi-Leduc flux was identified as the primary cause of the numerical instability since it depends on the magnetic field and thus can be either positive or negative. This necessitated modifications to the slope limiter to ensure the stability of the discretization scheme.

The model was validated through three test cases, with results demonstrating its ability to accurately predict the suppression and preheating effects induced by the electron nonlocal transport on both the perpendicular and Righi-Leduc heat flux, while also correctly characterizing suppression of the Biermann battery magnetic field and the shift of the peak position of the Nernst-generated electric field toward colder regions. The work advances the understanding of the electron transport in ICF and enhances the consistency between numerical simulations and experiments.

\section*{Acknowledgements}
\label{Acknowledgements}
This work was supported by the National Natural Science Foundation of China (Grant Nos. 12175309, 12475252 and 12275356), the Strategic Priority Research Program of Chinese Academy of Science (Grant Nos. XDA25050200 and XDA25010100), the Natural Science Foundation of Hunan Province, China (Grant No. 2025JJ20007), the Defense Industrial Technology Development Program (Grant. JCKYS2023212807) and the Postgraduate Scientific Research Innovation Project of National University of Defense Technology (XJJC2024041).

The software used in this work was developed in part by the DOE NNSA-and DOE Office of Science-supported Flash Center for Computational Science at the University of Chicago and the University of Rochester.

\bibliography{reference.bib}

\end{document}